\documentclass[pra,twocolumn,a4paper,groupedaddress,showpacs,floatfix,aps,10pt,
longbibliography,nobibnotes]{revtex4-1}
\usepackage{graphicx,amsmath,amssymb,amsfonts,dsfont,subfigure,color}

\usepackage[capitalise]{cleveref}
\usepackage{upgreek}

\newcommand{\mc}[1]{\ensuremath{\mathcal{#1}}}
\newcommand{\vroW}{\varrho}

\begin{document}

\title{Efficient microwave-to-optical conversion using Rydberg atoms}

\author{Thibault Vogt${}^{1,2}$}
\author{Christian Gross${}^{1}$}
\author{Jingshan Han${}^{1}$}
\author{Sambit~Bikas~Pal${}^{1}$}
\author{Mark Lam${}^{1}$}
\author{Martin Kiffner${}^{1,3}$}
\author{Wenhui Li${}^{1,4}$}

\affiliation{Centre for Quantum Technologies, National University of Singapore, 3 Science Drive 2, Singapore 117543${}^1$}
\affiliation{MajuLab, CNRS-UCA-SU-NUS-NTU International Joint Research Unit UMI 3654, Singapore 117543${}^2$}
\affiliation{Clarendon Laboratory, University of Oxford, Parks Road, Oxford OX1 3PU, United Kingdom${}^3$}
\affiliation{Department of Physics, National University of Singapore, Singapore 117542${}^4$}




\begin{abstract}

We demonstrate microwave-to-optical conversion using six-wave mixing in $^{87}$Rb atoms where the microwave field couples to two atomic Rydberg states, and propagates collinearly with the converted optical field. We achieve a photon conversion efficiency of $\sim 5$\% in the linear regime of the converter. In addition, we theoretically investigate all-resonant six-wave mixing and outline a realistic experimental scheme for reaching efficiencies greater than 60\%.

\end{abstract}

\maketitle

\smallskip
%

Rydberg atoms feature a quasi-continuum of narrow and strong dipole transitions coupling to microwave and terahertz (THz) radiation~\cite{gallagher:ryd}. Moreover, they can be easily excited and manipulated with laser light. These two properties endow Rydberg atoms with tremendous potential for applications combining optical light with microwaves or THz waves in precision spectroscopy, quantum sensing, and information processing.
For example, real-time THz field imaging has been achieved by THz induced optical fluorescence in Rydberg atoms~\cite{Wade2016}.
Sensitive detection of free-space microwave fields based on dressed electromagnetically induced transparency involving a Rydberg state (EIT) is being actively pursued~\cite{sedlacek2012microwave,simons2016using}.
Rydberg EIT can also be used to transfer digital information encoded in a microwave field onto an optical carrier, with applications in radio-over-fiber technologies~\cite{meyer2018digital,deb2018radio}.

Recently, microwave-to-optical conversion has been demonstrated employing frequency mixing via Rydberg states~\cite{han2018coherent}. The good coherence of the underlying process, characterized in Ref.~\cite{han2018coherent}, makes it a solid candidate for the transfer of quantum information between superconducting and photonic qubits.  Among the different technologies being developed for this application \cite{Hisatomi2016,Rueda2016,Bochmann2013,Bagci2014,higginbotham2018harnessing,PhysRevLett.119.063601}, the highest microwave-optical photon conversion efficiency $\eta$ obtained so far is nearly 50\%, but with a relatively narrow bandwidth of $\sim 12$~kHz~\cite{higginbotham2018harnessing}.
While the bandwidth of the converter using frequency mixing in Rydberg atoms demonstrated in ~\cite{han2018coherent} was as large as 4 MHz, $\eta$ was limited to 0.3\%. It thus remains to be shown that much higher conversion efficiencies can be achieved with this approach. 

In this letter, we demonstrate six-wave mixing in a cold rubidium (Rb) gas where all the waves are near-resonant with atomic transitions and propagate along a single axis.  In the first part of this letter, we describe the frequency-conversion mechanism, and present our results. We find that the response of our converter is linear for a wide range of microwave input powers. Most importantly, the chosen configuration, with collinear propagation of the waves, enables us to enhance the conversion efficiency by a factor of seventeen compared to that in~\cite{han2018coherent}. In the second part, we theoretically analyze all-resonant collinear six-wave mixing and identify the conditions for achieving high efficiencies.

\begin{figure}[htbp]
\centering
\includegraphics[width=0.85\linewidth]{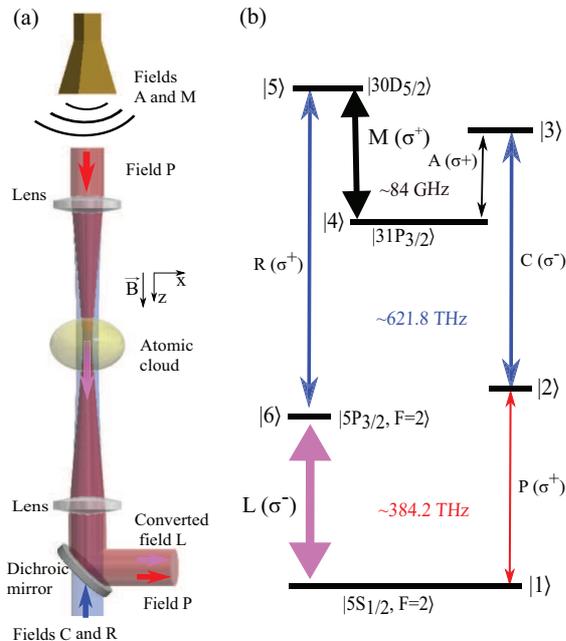}
\caption{(a) Experimental setup. The frequency mixing is performed with collinearly propagating fields in a Gaussian-distributed atomic sample. The laser beams are focused on the atomic ensemble, whereas the microwave fields are emitted from a horn antenna. (b) Energy level diagram and coupled transitions. The polarization of the fields are indicated inside parentheses.}
\label{Fig1}
\end{figure}

The principle of the experiment is as follows. A cloud of cold polarized $^{87}$Rb atoms is illuminated by four auxiliary electromagnetic fields as well as the microwave field M to be converted, as shown in Fig.~\ref{Fig1}(a). By non-linear frequency mixing of the six waves in the atomic medium, the field M is converted into the optical field L. The chosen configuration of energy levels is displayed in Fig.~\ref{Fig1}(b). The six waves P, C, A, M, R, and L are near-resonant with the atomic transitions shown in the figure, where $|1\rangle \equiv |5S_{1/2}, F = 2, m_F = 2\rangle$, $|2\rangle \equiv |5P_{3/2}, F = 3, m_F = 3\rangle$, $|3\rangle \equiv |30D_{3/2}, m_J = 1/2\rangle$, $|4\rangle \equiv |31P_{3/2}, m_J = -1/2\rangle$, $|5\rangle \equiv |30D_{5/2}, m_J = 1/2\rangle$, and $|6\rangle \equiv |5P_{3/2}, F = 2, m_F = 1\rangle$~\cite{note1}. In the absence of the microwave M, the system is in the configuration of microwave dressed Rydberg EIT formed by the two optical waves P and C, and the auxiliary microwave A. Once the M and R fields are added, the coherence induced between the ground state  $|1 \rangle$ and the intermediate state  $|6 \rangle$ triggers the generation of the converted optical field L.

\smallskip

In our setup, the fields P, M, and A are copropagating along the quantization axis $z$, and counterpropagating with the fields C and R.
The frequency differences between transitions $|1\rangle \leftrightarrow |2\rangle$ vs. $|1\rangle \leftrightarrow |6\rangle$, $|2\rangle \leftrightarrow |3\rangle$ vs. $|6\rangle \leftrightarrow |5\rangle$, and $|4\rangle \leftrightarrow |3\rangle$ vs. $|4\rangle \leftrightarrow |5\rangle$ are 0.27, 0.73, and 0.45 GHz, respectively, including the Zeeman shifts due to the 6.1~G bias magnetic field applied along the $z$ axis~\cite{ye1996hyperfine,mack2011measurement}. These differences are small compared to the frequencies of the fields, and to a good approximation their wavevectors obey $\mathbf{k}_C \approx \mathbf{k}_R$ and $\mathbf{k}_A \approx \mathbf{k}_M$, resulting in $\mathbf{k}_L \approx \mathbf{k}_P$ due to phase matching. Consequently, the L and P fields are copropagating, and all the fields are collinear. The P and L optical waves are of opposite polarization and are separated before simultaneous photodetection.

In more detail, the atomic cloud is Gaussian distributed, with $1/e^2$ radius $w_z=1.85$~mm, peak atomic density $n_{at,0}\sim 2 \times 10^{10} \; \textrm{cm}^{-3}$, and temperature $T\sim70 \; \mu$K. The optical beams P, C, and R are focused on the atomic cloud with beam waists $w_P$, $w_C$, and $w_R$ of 25, 54, and $45 \, \mu\textrm{m}$, respectively. The frequencies of the C and R fields are maintained on-resonance, and only that of the P field is varied to acquire spectra as a function of the frequency detuning $\Delta_P$. The microwaves are combined with a magic tee, and emitted  out of a horn antenna with linear transverse polarization. The  $\sigma^+$ polarization components of the microwaves form the M and A fields, and are kept on resonance with frequencies of $\nu_M=84.18 \; \textrm{GHz}$, and $\nu_A=83.72 \; \textrm{GHz}$, respectively. The $\sigma^-$ polarization components are off-resonant due to the Zeeman shifts, hence play a negligible role in this experiment.  The electric dipole matrix elements of the atomic transitions in Fig.~\ref{Fig1}(b) are, in Hartree atomic units, $|d_{21}|=2.99$, $|d_{61}|=1.22$, $|d_{32}|=0.00914$, $|d_{34}|=211$, $|d_{54}|=387$, and  $|d_{56}|=0.0138$.

\begin{figure}[tbp]
\centering
\includegraphics[width=\linewidth]{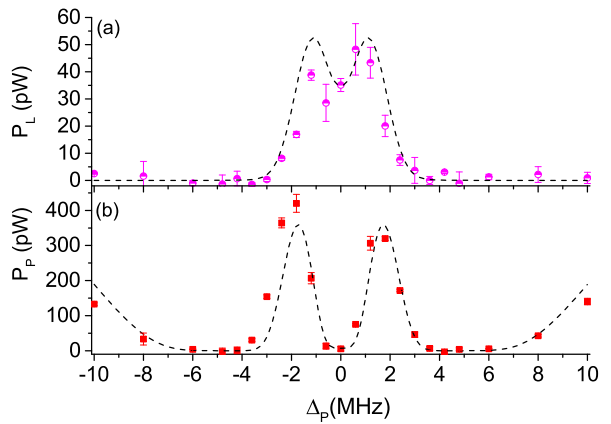}
\caption{Spectra of the generated light power  $P_L$ (a), and of the transmitted P field power $P_P$ (b). The dashed lines are simulated results obtained using Maxwell-Bloch's equations (see text).
}
\label{Fig2}
\end{figure}

Typical spectra of the measured powers $P_L$ and $P_P$ of the L and P fields vs. $\Delta_P$ are shown in Fig.~\ref{Fig2}. The peak Rabi frequencies of the incoming fields are $\Omega_{P0}/2 \pi=1.0 \; \textrm{MHz}$, $\Omega_{C0}/2 \pi=9.5 \; \textrm{MHz}$, $\Omega_{R0}/2 \pi=6.3 \; \textrm{MHz}$, $\Omega_{A0}/2 \pi=2.9 \; \textrm{MHz}$, and $\Omega_{M0}/2 \pi=1.4 \; \textrm{MHz}$. The data obtained for each $\Delta_P$ are averages of the signals recorded in steady state condition over $10\, \mu \textrm{s}$ after the application of the fields. The conversion is most efficient around $\Delta_P=0$ [Fig.~\ref{Fig2}(a)], which is consistent with the non-linearity responsible for the frequency mixing being maximum close to resonance. 
The spectra of the L and P fields in Fig.~\ref{Fig2} are approximately symmetric. The double-peak shape of the spectrum of $P_P$ occurs mostly due to the large A field, and is likely an effect of microwave dressed EIT on $|1\rangle \rightarrow |2\rangle \rightarrow |3\rangle \rightarrow |4\rangle$. 
We attribute the small reduction of $P_L$ at the center of the spectrum in Fig.~\ref{Fig2}(a) to the strong absorption of the P field around $\Delta_P=0$, which tends to reduce the effective volume of the medium where the conversion occurs.

\begin{figure}[htbp]
\centering
\includegraphics[width=\linewidth]{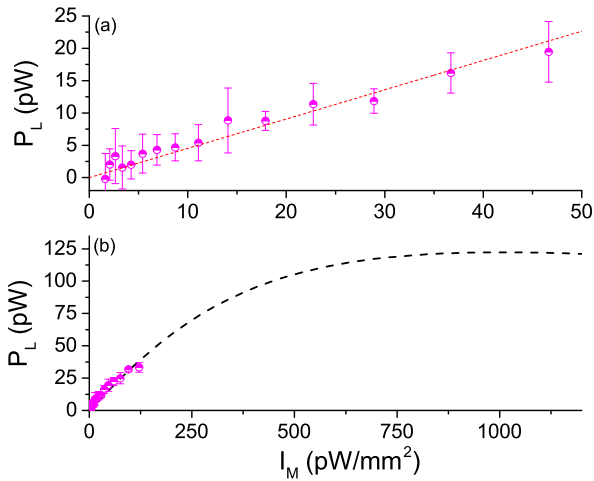}
\caption{The linearity of the converter. (a) The power of the generated light is plotted versus $I_M$ in the range of 0 to $50\ \textrm{pW/mm}^2$. The red dashed line is the result of a linear regression. (b) The power of the generated light is plotted versus $I_M$ in the range of 0 to $2000\ \textrm{pW/mm}^2$. The dashed line is the numerically simulated result (see text).}
\label{Fig3}
\end{figure}

The behavior of the measured power $P_L$ for $\Delta_P=0$ is approximately linear as a function of the input intensity $I_M$ of field M [see Fig.~\ref{Fig3}(a)]. Given $P_L \approx \alpha I_M$, a linear fit to the data yields the slope $\alpha = 0.45 \; \textrm{mm}^{2}$. The photon conversion efficiency of the process is deduced as 
\begin{equation}
\eta \approx \frac{P_L /\nu_L}{I_M S_M / \nu_M}=0.051,\label{eta}
\end{equation}
 where $S_M= \pi w_P^2$, and $\nu_L\approx 384.228 \; \textrm{THz}$. For this calculation, we consider the microwave photons incident on the conversion medium, whose transverse size is determined by the probe beam waist $w_P$. This conversion efficiency is seventeen times larger than the one reported in~\cite{han2018coherent} using a perpendicular configuration, where the microwave fields propagated at right angles to the optical fields. 
This is consistent with the reduction of $S_M$ by a factor of $\sim w_P/w_z$ due to the change of geometry. The physical reason for the enhancement is that the conversion occurs over a much longer distance in the collinear configuration ($\sim 2 w_z$) than that in the perpendicular geometry ($\sim 2 w_P$).

Next, we show that the interaction of the laser
and microwave fields with the atomic ensemble is well described within the framework of coupled
Maxwell-Bloch equations, which will allow us to theoretically investigate possible improvements to the present scheme. We restrict the problem to the calculation of one-dimensional solutions for the fields in the slowly varying envelope approximation, and in steady state. With these assumptions, the field amplitudes satisfy the following differential equations:

\begin{equation}
\partial_z  \Omega_X = i \frac{n_{\text{at}} |d_{ji}|^2 \nu_X}{h \, \epsilon_0 \, c} \vroW_{ji}. \label{maxwell}
\end{equation}
In this set of equations, the complex amplitudes $\mc{E}_{X}$ of each field X ($\text{X}\in\{\text{P},\text{R},\text{M},\text{C},\text{L},\text{A}\}$)  have been rescaled using the definition of the Rabi frequency $\Omega_{X}=2\mc{E}_{X} d_{ji} / \hbar$ where $\hbar$ is the reduced Planck constant and $d_{ji}$ is the dipole matrix element of the transition $|i \rangle \leftrightarrow |j \rangle$ driven by field X.
Moreover, $\vroW_{ji}$ is the corresponding atomic coherence, written in a suitable rotating frame removing fast temporal and spatial oscillations, $\nu_X$ is the frequency of field X, $n_{at}$ is the atomic density, $\epsilon_0$ is the free-space permittivity, and $c$ is the speed of light in vacuum. 
The atomic coherences are given by the steady state solution of the following Markovian master equation
\begin{equation}
\partial_t \vroW = - \frac{i}{\hbar} [ H , \vroW ]
+\mc{L}_{\Gamma}\vroW+\mc{L}_{\text{deph}}\vroW\,,
\label{master_eq}
\end{equation}
where $H=- \hbar \left(\Delta_P \, |2\rangle \langle 2|+\sum_X \Omega_X |j\rangle \langle i| + \textrm{H.c.} \right)/2 $ is the interaction Hamiltonian between a single atom and the fields. The Lindblad term $\mc{L}_{\Gamma}\vroW$ accounts for spontaneous emission from the intermediate states $|2\rangle$ and $|6\rangle$ with rate $\Gamma=2 \pi \times 6.067 \; \textrm{MHz}$ and from the Rydberg states. The term
$\mc{L}_{\text{deph}}\vroW$ accounts for additional dephasing mechanisms, and may be written as 

\begin{equation}
\mc{L}_{\text{deph}}\vroW = \sum_{j \in \{ 3, 4, 5\}} \gamma_{j} \left( 2 P_{j} \vroW P_{j} - P_{j} \vroW -\vroW P_{j}  \right),
\label{Dissipator}
\end{equation}
where $P_{j}$ is the projection operator on state $|j\rangle$, and $\gamma_j$ is the dephasing rate associated with state $|j\rangle$.
These rates include the effects of atomic collisions, dipole-dipole interactions
between Rydberg atoms, and finite laser linewidths. The dephasing rates $\gamma_{1}$, $\gamma_{2}$, and $\gamma_{6}$ are neglected since they are much smaller than $\Gamma$.
The solutions of the system of \cref{maxwell,master_eq} are integrated along the transverse direction to take into account the finite size of the beams, while neglecting any variation of $\Omega_C$ and $\Omega_R$ along $z$ since $|d_{32}|$ and $|d_{56}|$ are very small. The simulation results [see Figs. \ref{Fig2}(a), \ref{Fig2}(b), and \ref{Fig3}(b)] are in good agreement with the data for $\gamma_{3} = 2 \pi \times $50 kHz,  $\gamma_{4} = 2 \pi\times$ 300 kHz, and $\gamma_{5} = 2 \pi \times$ 400 kHz, while the other input parameters are obtained from experimental calibrations. This further asserts the validity of the model of Refs.~\cite{kiffner2016,han2018coherent} and allows us to extrapolate the behavior of the data to larger input microwave intensities, as shown in Fig.~\ref{Fig3}(b). 
While $P_L$ is linear at small $I_M$, it starts to saturate when $\Omega_{M0}$ becomes of the order of $\Omega_{A0}$.

To present a pathway to improving the efficiency, we will take a closer look at \eqref{master_eq}, in the case of all-resonant waves. The analytical steady-state solution of \eqref{master_eq} is obtained by an expansion of $\varrho$ up to third order in the weak fields $\Omega_P$, $\Omega_M$, $\Omega_L$, and the small dephasing rates $\gamma_j$~\cite{kiffner2016}. In keeping only the two leading order terms, the coherences $\varrho_{21}$, $\varrho_{61}$, and $\varrho_{54}$ take the following simple form~\cite{note2}
\begin{align}
\varrho_{21} & =  i \frac{\Omega_P}{\Gamma}, \label{rho21}\\
\varrho_{61} & = i \frac{\Omega_P \Omega_C \Omega_M }{\Gamma \Omega_A \Omega_R},\label{rho61}\\
\varrho_{54} & = -i \frac{\Omega_R \Omega_L \Omega_A \Omega_C^* \Omega_P^*-|\Omega_C|^2  |\Omega_P|^2 \Omega_M }{\Gamma |\Omega_A|^2 \; |\Omega_R|^2 }, \label{rho54}
\end{align}
where the terms proportional to the dephasing rates have been omitted. These latter terms are comparatively very small if the conditions $\Gamma |\Omega_A|^2 /|\Omega_C|^2 \gg \gamma_4 $, $|\Omega_R|^2/ \Gamma \gg \gamma_5$, and $|\Omega_C \Omega_R \Omega_M \Omega_P | / |\Gamma \Omega_A \Omega_L  | \gg \gamma_5$ are fulfilled. For finite dephasing rates as measured in our current experiment, these conditions are best met experimentally if $|\Omega_A|$ is comparable to $|\Omega_C|$ and $|\Omega_R| \gg \Gamma$.  Having obtained a solution approximately insensitive to $\gamma_{3,4,5}$ is important as we seek a system that is robust against interaction-induced dephasing to realistically reach high efficiencies. In this regime, and taking the large Rabi frequencies $\Omega_A$, $\Omega_C$, and $\Omega_R$ as constants,  we find from the system of \cref{rho21,rho61,rho54,maxwell} that $\Omega_L$ satisfies the following differential equation
\begin{equation}
 \partial^2_v  \Omega_L + a_2 \left( 1 - v \right)  \partial_v \Omega_L +  a_1^2 \,  \Omega_L = 0,\label{EquaDif}
\end{equation}
where $v$ is defined as $ v(z)=1-e^{-u(z)/2}$, with $u(z) = \int_{-\infty}^z \textrm{d}z' \, 2 n_{at} |d_{21}|^2 \nu_P / \left( h \epsilon_0 c \Gamma \right)$ being the optical depth vs. $z$ for the P field. Taking $\nu_P \approx \nu_L$, we have $a_1=c_1 |d_{61}/d_{21}|^2$ and $a_2=c_1^2 |d_{61}/d_{21}|^2$, where
\begin{equation}
c_1= \, \left|\frac{d_{54} }{d_{61}}\right| \sqrt{ \frac{\nu_M}{\nu_L }} \frac{\Omega_{C} \, \Omega_{P}\left(- \infty\right)}{ \Omega_{R} \, \Omega_{A} }. \label{a1}
\end{equation}
The analytical solution to \eqref{EquaDif} can be expressed in terms of a hypergeometric function and a Hermite polynomial.  \eqref{EquaDif} is very similar to that of a damped harmonic oscillator, with the damping term weakly depending on $v$, as $a_2 (1-v)$, where $0 \leq v < 1$. By analogy, when $a_2 \ll a_1$ the solution takes the simple form $\Omega_L(v) \approx \Omega_{L0} \sin \left( a_1 \, v \right)$, where $\Omega_{L0}=\sqrt{\nu_L / \nu_M} |d_{61}/d_{54}|\Omega_{M}\left(- \infty\right)$ corresponds to 100\% conversion efficiency. Therefore, optimizing the efficiency implies to fulfill the two conditions, $a_1 v(+\infty)= \pi/2$ and $a_2 \ll a_1$. For a large enough atomic cloud, this can be achieved by decreasing $|d_{21}|$, and increasing $|d_{61}|$. Physically, the idea is to minimize the absorption of the P field, while maximizing the converted L field. 

\begin{figure}[htpb]
\centering
\includegraphics[width=0.9\linewidth]{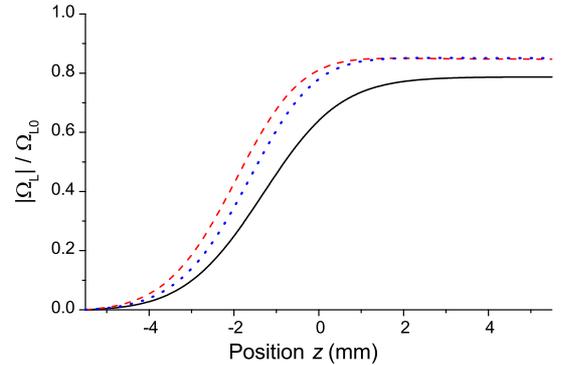}
\caption{Optimizing the converted field L. (Dashed red line) Analytical solution of the L field Rabi frequency $\Omega_L$ versus position $z$ inside a Gaussian-distributed atomic sample. The calculation is based on the result of \eqref{EquaDif}, with $a_1\, v(+\infty)= \pi/2$ (see text). (Blue dotted line) Full numerical simulation of $\Omega_L$ versus $z$, without considering dephasing rates. (Black solid line) Full numerical simulation of $\Omega_L$ versus $z$, including the dephasing rates $\gamma_3$, $\gamma_4$, and $\gamma_5$. For all three curves, $\Omega_L$ is rescaled with the Rabi frequency $\Omega_{L0}$ corresponding to 100\% efficiency.}
\label{Fig4}
\end{figure}

Based on this analytical derivation, we select a more favorable configuration of energy levels where $|d_{21}|$ is reduced by a factor of $\sqrt{6}$, and $|d_{61}|$ increased by $\sqrt{6}$. The new configuration is such that $|1'\rangle \equiv |1 \rangle$, $|2'\rangle \equiv |5P_{3/2}, F = 2, m_F = 1\rangle$, $|3'\rangle \equiv |3\rangle$, $|4'\rangle \equiv |31P_{3/2}, m_J = 3/2\rangle$, $|5'\rangle \equiv |30D_{5/2}, m_J = 5/2\rangle$, and $|6'\rangle \equiv |5P_{3/2}, F = 3, m_F = 3\rangle$, and we keep the fields as denominated in Fig.~\ref{Fig1}, but with modified polarizations and strengths. We consider extended input fields of Rabi frequencies $\Omega_{P0}/2 \pi=1.0 \; \textrm{MHz}$, $\Omega_{C0}/2 \pi=13.5 \; \textrm{MHz}$, $\Omega_{R0}/2 \pi=29 \; \textrm{MHz}$, $\Omega_{A0}/2 \pi=5.5 \; \textrm{MHz}$, and a Gaussian-distributed atomic cloud with $n_{at,0}=2 \times 10^{10} \; \textrm{cm}^{-3}$ and $w_z=3.7$~mm, such that $a_1 \, v(+\infty)\approx \pi/2$. Figure~\ref{Fig4} compares the analytical solution of \eqref{EquaDif} to the numerical simulation of $\Omega_L$ given by \cref{maxwell,master_eq} without dephasing, as a function of the position $z$ along the atomic cloud.  The two curves are in good agreement and the remaining discrepancy would vanish for $\Omega_{P0}/2 \pi <0.5 \; \textrm{MHz}$. Moreover, the introduction of finite dephasing rates similar to the ones measured in our experiment (see \eqref{Dissipator}) affects only moderately the numerical simulation, as expected from our initial assumptions and shown in Fig.~\ref{Fig4} as well. The efficiency $\eta$, deduced directly from these solutions, reaches close to 62\% if the dephasing rates are taken into account, in comparison to about 72\% without dephasing. This system would likely require a larger bias magnetic field and active optical pumping to compensate for the depumping effect of the $\sigma^-$ P field, which drives an open transition.
A possible approach to further improving $\eta$ is to rely on a higher lying state $|2\rangle$ such that $|d_{21}|$ is significantly reduced.

In conclusion, we have experimentally demonstrated efficient microwave-to-optical conversion using all-resonant six-wave mixing via Rydberg states. We have theoretically analyzed the conversion process in steady state, and identified a clear strategy for improving the all-resonant scheme. To reach unit-conversion efficiency for quantum applications, one may consider stimulated Raman adiabatic passage, a technique which may be applicable to our system~\cite{gard2017microwave}. Another option that in theory achieves near-unit conversion efficiency is to consider tuning two of the fields off-resonance to realize an effective two-photon transition, for example fields C and A in our system~\cite{kiffner2016,vogt2018microwave}. Eventually, the realization of microwave-to-optical conversion at the single photon level for sensing or quantum applications will require one to tightly focus or confine the microwave field, and integrate the system in a noise-free environment~\cite{Hogan2012,hermann2014long,cano2011experimental}.

\section*{Acknowledgments}

The authors acknowledge the
support by the National Research Foundation, Prime
Minister's Office, Singapore and the Ministry of Education, Singapore under the
Research Centres of Excellence programme. This work is supported by Singapore
Ministry of Education Academic Research Fund Tier 2 (Grant No.
MOE2015-T2-1-085).


\begin{thebibliography}{10}
\bibitem [{note1()}]{note1}%
  \BibitemOpen
  \bibinfo {note} {Note that the hyperfine structure of the considered Rydberg states is negligible.}\BibitemShut {Stop}%
\bibitem [{note2()}]{note2}%
  \BibitemOpen
  \bibinfo {note} {We have neglected the second leading order terms proportional to $\Omega_L$ in \cref{rho21,rho61,rho54}, as they are substantially smaller. The coherence $\varrho_{43}$ is of third order and proportional to $\Omega_L^*$, hence $\rho_{43}\approx 0$.}\BibitemShut {Stop}%
\end{thebibliography}
%

\end{document}